\documentclass [aps,pra,amssymb,amsmath,showpacs,twocolumn]{revtex4-1}
\usepackage[latin9]{inputenc}
\usepackage{amsmath}
\usepackage{amssymb}
\usepackage{graphicx}
\usepackage{verbatim}
\usepackage{bm}
\usepackage{color}

\usepackage{graphicx,epsfig,amsfonts,amssymb}
\usepackage{bm}% bold math
\usepackage{times}
\usepackage{lipsum}
\usepackage{verbatim}

\newcommand{\nn}{\nonumber}

\newcommand{\la}{\langle}
\newcommand{\ra}{\rangle}
\newcommand{\rar}{\rightarrow}

\newcommand{\be}{\begin{eqnarray}}
\newcommand{\ee}{\end{eqnarray}}

\newcommand{\bs}{\begin{equation}\begin{split}}
\newcommand{\es}{\end{split}\end{equation}}

\newcommand{\sn}{\operatorname{sn}}
\newcommand{\cn}{\operatorname{cn}}
\newcommand{\dn}{\operatorname{dn}}

\newcommand{\sech}{\operatorname{sech}}

\date{\today}

\begin{document}
\title{Nearly-optimal quantum control: analytical approach}

\author {Chen Sun$^{a,b}$, Avadh Saxena$^{a,c}$, and Nikolai A. {Sinitsyn}$^{a}$ }
\affiliation{$^a$ Theoretical Division, Los Alamos National Laboratory, Los Alamos, NM 87545,  USA}
\affiliation{$^b$ Department of Physics, Texas A\&M University, TX 77840,  USA}
\affiliation{$^c$ Center for Nonlinear Studies, Los Alamos National Laboratory, Los Alamos, NM 87545, USA}

\begin{abstract}
We propose nearly-optimal control strategies for changing states of a quantum system. We argue that quantum control optimization can be studied analytically within some protocol families that depend on a small  set of parameters for optimization. This optimization strategy can be preferred in practice because  it is physically transparent and does not lead to combinatorial complexity in multistate problems.
For demonstration, we design optimized control protocols that achieve switching between orthogonal states of a naturally biased quantum two-level system.
\end{abstract}

\maketitle

%\section{Outline}
%1. introduction

%2. model 1, infinite sweep, spin flip

%3. model 2, infinite sweep, spin rotation

%. infinite time case -- solitons

%. finite time case -- elliptic functions

\section{Introduction}

The future miniaturization technologies will  face with the challenge to design elementary information-processing units that  employ quantum effects and operate  with simultaneously  high speed and low dissipation. Such requirements are usually  contradictory to each other. Therefore, they raise the need for quantum optimal control, i.e., the design of protocols for application of time-dependent fields that lead to the desired behavior of a quantum system while minimizing some cost function \cite{grigorenko-book}.

There are two major complications with control optimization at the quantum  level.
 First, such a control has to be  nonlinear. This follows already from the fact that  physical characteristics of a quantum system are nonlinear functions of dynamic variables, which are the state vector components. Moreover, control fields multiply the state vector in the Schr\"odinger equation, so this equation is nonlinear if we treat both the state vector and control fields as dynamic variables. {Even for a spin-1/2 system, conventional optimal control already becomes complex, requiring involved mathematical treatment and sometimes numerical calculations to find the optimal protocols \cite{Boscain1,Dalessandro,Boscain2,Garon}.} The second complication is the exponentially fast growth of the number of variables that parametrize the state vector. For example, $N$ qubits are described quantum mechanically by a state vector with $2^N$ components. So, beyond the most elementary cases, even numerical solution of a quantum optimization problem  by conventional methods of the control theory is hard to achieve.

 In this article, we argue that complexity of the optimal quantum control problem can  be considerably reduced in many applications. We propose to search for the desired optimization only within some families of  control protocols that allow analytical solution of the nonstationary Sch\"odinger equation. If such a family is sufficiently broad, one can expect that optimization within this family will lead to a reasonable cost while the problem will be tractable.

 Finding large classes of exactly solvable models with desired properties may look impossible at first view. Indeed,  there is even no known general analytical solution for the dynamics of a spin-1/2 system in an arbitrary time-dependent magnetic field. Higher dimensional cases are even more complex because they require solutions of higher than 2nd order differential equations with time-dependent parameters, which remain poorly understood.

 Contrary to such expectations, we would like to point out that it is actually not hard to generate families of time-dependent control protocols whose effects on quantum systems can be understood analytically. One possibility to do this is  to apply control fields that simply compensate for terms in the Hamiltonian that are responsible for complex behavior. For example, if there are controlled couplings between some qubits, we can simply set such couplings to zero in order to make individual qubit dynamics easy to control by simple local fields for a while. It is then possible to determine and optimize costs within the family of such protocols analytically.

Another possibility to design a family of solvable Shr\"odinger equations is to solve an inverse problem. Namely, in equation
\be
i\frac{d\Psi}{dt} = \hat{H}(t) \Psi,
\label{shrod}
\ee
we can prescribe the functional form for some desired types of dynamics of $\Psi(t)$ and then treat (\ref{shrod}) as the linear equation for unknown elements of the matrix Hamiltonian $\hat{H}(t)$. Since the number of components of the state vector scales linearly with the size of the phase space $N$ and the number of components of the matrix $\hat{H}$ scales as $N^2$, this equation can generally be solved with large redundancy. Many families of such obtained exactly solvable models are  known \cite{garanin,sarma,berry}. Models generated this way describe somewhat unnatural dynamics but this is not a drawback for quantum control purposes.
After parametrizing the class of solutions of such an inverse problem, we can explore the cost function within the resulting parameter space.

In the rest of the article, we demonstrate these two strategies using the model of a spin-1/2 system control by a time-dependent magnetic field. {Instead of straightforward global optimization, we are going to use the fact that there are classes of protocols for which analytical solutions of the Schrodinger equation are known. Within such classes the optimization problem strongly simplifies, but generally the result will not coincide with a fully optimized solution. In this sense, our result is ``nearly-optimal''.}

 %Formal approach to the quantum control  \cite{grigorenko}  would based on minimization of a specific Lagrangian that includes terms responsible for the cost function supplemented by a constraint describing the evolution of the density matrix. Introduction of such a constraint requires doubling of variables, i.e., supplementing all physical variables  by Lagrange multiplies. For a spin-1/2, the functional for minimization  will generally depend on nine dynamic variables that include three variables to parametrize the density matrix, their Lagrange multipliers, and three components of the  magnetic field for control.  Corresponding set of nonlinear coupled differential equations, for the case with $\varepsilon \ne 0$,  can be solved only numerically.
 %In sections 2 and 3 we discuss two particularly simple examplary problems that we solve for optimal control protocols in terms of elementary functions. In section 4, we consider complications that emerge from more realistic constraints on the control protocols.  We summarize our findings and discuss the perspectives in section 5.

\section{The model}
\label{model}
\begin{figure}[!htb]
\scalebox{0.55}[0.55]{\includegraphics{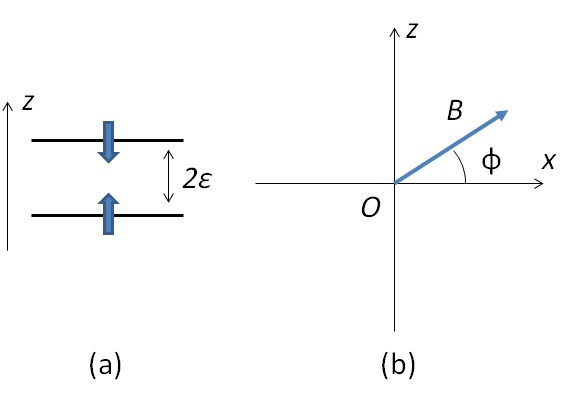}}
%\hspace{-2mm}\vspace{-4mm}%!!
\caption{(a) The two-level system with intrinsic energy splitting $2\varepsilon$. (b) The effective field in the $xz$-plane, which is characterized by size of the field $B$ and the angle $\phi$.}
\label{levels}
\end{figure}

Consider a two level system shown in Fig.~\ref{levels}(a). We assume that the  system is initially in the ground state that is separated from the excited state by a natural energy splitting bias $2\varepsilon$. So, the free Hamiltonian of the system is
\begin{align}\label{}
\hat{H}^{\rm free}=-\varepsilon\hat{\sigma}_z,
\end{align}
where $\hat{\sigma}_z$ is the Pauli operator. Assume that the goal is to design pulses of some control field that will lead to the definite transition from the ground state to the excited state.
It is convenient to think about a two-level system in terms of a spin-1/2 in a time-dependent magnetic field. So, the general Hamiltonan is
\begin{align}
\label{magn}
\hat{H}=-\varepsilon\hat{\sigma}_z+B_x^{\rm ex}(t)\hat{\sigma}_x +B_y^{\rm ex}(t)\hat{\sigma}_y +B_z^{\rm ex}(t)\hat{\sigma}_z,
\end{align}
where $B_x^{\rm ex}$, $B_y^{\rm ex}$, and $B_z^{\rm ex}$ are control parameters, which we will call components of the external magnetic field.
There are numerous known ways to design pulses of a magnetic field in order to flip the spin. However, practical situation may impose conditions that favor some protocols over the others. Specifically, we will assume that following four conditions must be met:

(i) During the spin-flip, nonzero average spin polarization along  $y$-axis should be avoided by all means possible. Thus we require that $\la \Psi(t)| \hat{\sigma}_y |\Psi(t) \ra = 0$ at any time.

(ii) We want to minimize time that the spin spends not being strictly polarized along $z$-axis. In other words, we want to flip the spin as fast as possible.

(iii) The size of the external control field that we apply should be minimized.

(iv) We are looking for protocols whose properties can be written in terms of some known special functions.

Here we would like to stress that our choice of conditions (i)-(iv) is not motivated by some immediate contemporary experimental research. We chose them  for purely illustrative reasons because they provide an example that is sufficiently nontrivial mathematically to illustrate advantages and possible weaknesses of our approach. 

For definiteness, we will  assume that the control protocol starts at some negative time moment so that spin appears in the state ``up" by time $t=0$, after which all external fields are switched off.
Condition (i) may be justified in real situation by requirement to avoid unwanted coupling with ambient qubits that are placed along the $y$-axis. We assume that this condition is not negotiable so it merely restricts the allowed phase space for dynamics. In particular, this condition forbids application of popular Rabi-like pulses that induce circulation of spin polarization in the $xy$-plane.  

Condition (iv)  expresses the desire to work with pulse shapes with easily characterizable properties. This means that we are not looking for a complete solution
of the optimization problem in terms of some nonlinear differential equation that can be solved only numerically. Instead, we restrict our search to a specific class of protocols with known analytical solutions. This class is quite large so we hope that the deviation of the cost of our protocol is not essentially different from the cost of a numerically exact solution of the optimization problem.

Conditions (ii) and (iii) are conflicting with each other. In order to minimize time of spin flip, we should apply strong control fields but such fields induce unwanted dissipation that we also want to minimize. In order to resolve this conflict, we should
quantify  conditions (ii) and (iii) by introducing some  cost functional $C$.  We will assume the simplest form of $C$ that is consistent with symmetry of the problem and condition that this functional should involve not higher than 2nd order powers of expectations of Pauli matrices and control field components:
\begin{widetext}
\begin{equation}
\label{new_cost_def_2}
C=\int_{-\infty}^{0}dt \, \left\{ A
 \la \Psi(t) | \hat{\sigma}_x | \Psi(t) \ra ^2 +(B_x^{\rm ex}(t))^2+(B_y^{\rm ex}(t))^2  +
(B_z^{\rm ex}(t))^2 \right\}.
\end{equation}
\end{widetext}
Here the first term inside the integral penalizes all values of the state vector $|\Psi(t) \ra$ that have nonzero expectation value of $\hat{\sigma}_x$ operator. Since we are interested in protocols that asymptotically have  $\la \Psi(t)| \hat{\sigma}_x |\Psi(t) \ra^2 = 0$ at $t \rar - \infty,0$, the time integral of this expression has the meaning of the effective time of spin flip.
If the external field were the true magnetic field, the physical meaning of the rest of the cost would be merely the energy of the pulse. So, naturally, this cost favors application of smaller magnetic field values.
Parameter $A$ describes the relative importance of the two cost contributions. Bigger values of $A$ favor faster protocols that require larger external field amplitudes.  We will not consider other possible restrictions that have been encountered in the literature \cite{other}.

We can now quantify the problem. Our goal  is to find the time-dependent external field that induces dynamics of the state vector that minimizes the functional of the form
\begin{align}\label{new_cost_2}
&C=\int_{-\infty}^{0}\mathcal{L} dt,
\end{align}
 under condition that  during the evolution we have $\la \Psi(t) | \hat{\sigma}_y | \Psi(t) \ra =0$ and the spin is fully polarized in opposite directions along $z$-axis at the beginning and the end of the protocol.

\section{Optimal square pulse protocol}
Without the natural bias (i.e., at $\varepsilon=0$), we would be able to rotate the spin by merely applying the external field along the $y$-axis. This problem is exactly solvable and its optimization was solved in Ref.~\cite{grigorenko}. At $\varepsilon \ne 0$, the formal approach to optimization, however, would involve introduction of Lagrange multipliers that account for evolution of the state vector and constraints.
In contrast, as our first strategy, we consider a straightforward way to rotate the spin while keeping its polarization within $xz$-plane: we just apply an external field in the $z$-direction, i.e., momentarily set $B_z^{\rm ex}=\varepsilon$, in order to compensate for the intrinsic field bias.
Simultaneously, we apply a constant field along the $y$-direction while keeping $B_x^{\rm ex}=0$. As a result, the spin will precess around $y$-axis.
Let $\phi$ be the angle that our spin makes with the $x$-axis, as shown in Fig.~\ref{levels}(b).
For precession in this constant magnetic field we have
\be
\la \Psi(t) | \hat{\sigma}_x | \Psi(t) \ra^2=\sin^2(2B_y^{\rm ex} t).
\label{free}
\ee
We should switch the external field off  when the spin completes rotation by an angle $\pi$.
Obviously, for uniform rotation, our protocol should have a time duration
\be
T_u=\pi/(2B_y^{\rm ext}),
\label{time-free}
\ee
where the subscript ``u" stands for ``uniform", and the cost is
\begin{align}\label{}
&C_{\rm u}=\int_{-T_u}^0 \mathcal{L}dt=\frac{\pi^2}{4T_u}+\left(\varepsilon^2+\frac{A}{2}\right)T_u.
\end{align}
Using (\ref{time-free}), this cost is minimized at $B_{y}^{\rm ex}=\sqrt{\varepsilon^2+\frac{A}{2}}$:
\begin{align}\label{result_URP}
&C_{u}^{\rm min}=\pi\sqrt{\varepsilon^2+\frac{A}{2}}.
\end{align}

The square-pulse protocol is simple but it is expected to be strongly suboptimal because minimization is performed only within a family of field pulses parametrized by a single parameter $T_u$.
In what follows, we are going to explore more complex strategies. First, we can  improve the above strategy by allowing a more complex time-dependence of $B_y^{\rm ex}(t)$.   Second, we can consider a different family of protocols that have nonzero values of $B_x^{\rm ex}(t)$.

\section{Protocols with time-dependent $B_y^{\rm ex}$}
\label{sine-G}

Let us now assume that we again apply the field with $B^{\rm ex}_z=\varepsilon$ and $B^{\rm ex}_x=0$ during some time $T_0$.  Now, we treat $T_0$ as a free parameter and we also allow arbitrary time-dependence of $B_y^{\rm ex}(t)$. The only constraint on the transverse field  is that
$$
2\int_{-T_0}^0 B_y^{\rm ex}(t) \, dt = \pi.
$$
We have then
\be
\la \Psi(t) | \hat{\sigma}_x | \Psi(t) \ra^2=\sin^2 \left(\int_{-T_0}^{t} 2B_y^{\rm ex} (t') \, dt' \right).
\label{free2}
\ee
To simplify this expression, we will introduce the phase variable
$$
\phi(t) \equiv -\pi/2+2 \int_{-T_0}^{t} B_y^{\rm ex} (t') \, dt' ,
$$
in terms of which the cost functional now reads
\begin{align}\label{cost1}
%&C_{sG}[\phi(t)]=\varepsilon^2 T_0 +\int_{-T_0}^0 \left\{ \frac{\dot{\phi}^2}{4} +A\cos^2(\phi) \right\}dt, \nn\\
&C_{sG}[\phi(t)]=\int_{-T_0}^0 \left\{ \frac{\dot{\phi}^2}{4} +\varepsilon^2+A\cos^2(\phi) \right\}dt, \nn\\
& \phi(-T_0)=-\pi/2, \quad \phi(0) = \pi/2,
\end{align}
where the dot denotes differentiation with respect to $t$ and the subscript ``$sG$" stands for ``sine-Gordon" because Eq.~(\ref{cost1}) coincides with a Langrangian of a classical mechanical motion of a particle along coordinate $\phi$ in the sine-Gordon potential $-A\cos^2(\phi) $. The  term with $\varepsilon^2$ in (\ref{cost1}) should be kept because it penalizes protocols with larger duration $T_0$.

Varying the cost functional over $\phi(t)$ one can find the equation of motion. Instead of solving this equation, we invoke the classical mechanical analogy and say that the energy $e$ of the particle motion is conserved:
\begin{align}\label{}
\frac{1}{4}\dot{\phi}^2-\varepsilon^2-A\cos^2\phi=e = {\rm const}.
\end{align}
Hence,
%This right region is finite and $E$ is not necessarily zero. We will then treat $E$ to be a parameter, with respect to which we minimize the cost in this region. From Eq. \eqref{L_right} we have
\begin{align}\label{phi-AE}
\dot{\phi}=2\sqrt{\varepsilon^2+A\cos^2\phi+e}.
\end{align}

Formally, duration of the protocol $T_0$ should be also optimized. However, varying cost over this parameter we find the value of energy that cannot be used to satisfy the boundary conditions.
Instead, we will treat $e$ as an
additional parameter for optimization.
Integrating over time we find (definitions of special functions are provided in appendix A):
\begin{align}\label{}
&\phi=\arcsin\left[\sn\left(2\sqrt{\varepsilon^2+A+e}(t+T_0)|\frac{A}{\varepsilon^2+A+e}\right)\right],
\end{align}
and using the boundary conditions, $\phi(t=-T_0)=-\pi/2$ and $\phi(t=0)=\pi/2$, we obtain
\begin{align}\label{relation_AaT2}
T_0(e)=\frac{1}{\sqrt{\varepsilon^2+A+e}}K\left(\frac{A}{\varepsilon^2+A+e}\right),
\end{align}
where $K(m)$ is the complete elliptic integral of the first kind (appendix A). %, defined as

Substituting (\ref{phi-AE})-(\ref{relation_AaT2}) into (\ref{cost1}) we find
\begin{align}\label{cost1-sG}
&C_{ sG}=-eT_0(e)+ 2\sqrt{\varepsilon^2+A+e}E\left(\frac{A}{\varepsilon^2+A+e}\right),
\end{align}
where $E(m)$ is the complete elliptic integral of the second kind. %defined as
%\begin{align}\label{def_E_com}
%&E(m)=\int_0^{\frac{\pi}{2}}d\theta\sqrt{1-m\sin^2\theta}.
%\end{align}
%Differentiating $C_{i}$ over $E$, we find that the derivative is zero only at $E=0$. One can check that the $E=0$ corresponds to a minimum.
Finally, we should minimize the cost over the remaining free parameter $e$. Differentiating  (\ref{cost1-sG}) with respect to $e$, we find that the minimum is achieved at $e=0$ so,
\begin{align}\label{}
&\phi(t)=\arcsin\left[\sn\left(2\sqrt{\varepsilon^2+A}(t+T_0)|\frac{A}{\varepsilon^2+A}\right)\right], \nn\\
& \pi/2\ge \phi \ge-\pi/2,
\end{align}
where $T_0$ is given by
\begin{align}\label{}
T_0=\frac{1}{\sqrt{\varepsilon^2+A}}K\left(\frac{A}{\varepsilon^2+A}\right),
\end{align}
and  the minimal cost within this family of protocols is given by
\begin{align}
\label{costG}
C_{sG}^{\rm min}=2\sqrt{\varepsilon^2+A}E\left(\frac{A}{\varepsilon^2+A}\right).
\end{align}

\section{Optimal shortcut to adiabaticity}
\label{shortcut}
A bigger class of analytically tractable protocols can be obtained by  solving the inverse problem to the nonstationary Schr\"odinger equation. Many techniques to do this have been developed for spin-1/2 systems \cite{garanin,sarma,berry}. Here we will
focus on the class of such solutions called shortcuts to adiabaticity \cite{berry,Demirplak_2008,Deffner_2014,Bason_2011,Malossi_2013,Zheng_2016}. This class covers not only spin-1/2 but also many practically interesting
 multistate situations \cite{delCampo}.

Let $\hat{H}_0(t)$ be some time-dependent Hamiltonian and  $|u(t)\ra$ be one of the instantaneous eigenstates of this Hamiltonian, which continuously depends on $t$.  Note that $|u(t)\ra$ is generally not a solution of the nonstationary Schr\"odinger equation with $\hat{H}_0(t)$.  The Hamiltonian of a  shortcut to adiabaticity is  obtained by adding a counterterm $\hat{H}_{ct}(t)$ such that the evolution of the state vector with the Hamiltonian $\hat{H}(t)=\hat{H}_0(t) +\hat{H}_{ct}(t)$ does follow the path of eigenstates $|u(t)\ra$. It turns out that there is a formal expression for such a counterterm \cite{berry,Bason_2011,Zheng_2016}:
  \be
 \hat{H}_{ct}=\frac{i}{2}\left(| \partial_t  u(t)\ra \la u(t) | -|u(t)\ra \la \partial_t u(t) | \right).
 \label{ct1}
 \ee
In other words, the instantaneous eigenvector $|u(t)\ra$ of  $\hat{H}_{0}(t)$ becomes the exact solution of the nonstationary Schr\"odinger equation (\ref{shrod})  with the Hamiltonian $\hat{H}(t)=\hat{H}_0(t)+\hat{H}_{ct}(t)$, where the counterterm is given by (\ref{ct1}).

For the following discussion, it will be convenient to introduce a new vector  ${\mathbf B}$ with absolute value $B$ and components
$
B_x\equiv B\cos(\phi) = B_x^{\rm ext};\,\,\, B_y=0; \,\,\, B_z \equiv B\sin(\phi) = B_z^{\rm ext}-\varepsilon.
\label{bdef}
$
If now we choose
\begin{align}\label{}
\hat{H}_0(t)=B_z(t)\hat{\sigma}_z+B_x (t)\hat{\sigma}_x,
\end{align}
then, as desired, we will have evolution of the eigenstate $|  u(t)\ra$ such that spin polarization will be always directed along the field ${\mathbf B}$, which is restricted to $xz$-plane.
For such a  spin-1/2 Hamiltonian, the counterterm is
 \cite{Bason_2011,Malossi_2013}: %of the form[]:
\begin{align}\label{}
&\hat{H}_{ct}(t)=\frac{1}{2}\frac{d}{dt}\left(\arctan\frac{B_x}{B_z}\right)\hat{\sigma}_y=-\frac{\dot{B}_zB_x-B_z\dot{B}_x}{2\left[(B_z)^2+(B_x)^2\right]}\hat{\sigma}_y,
\end{align}
which can also be expressed as
\begin{align}\label{}
\hat{H}_{ct}=-\frac{1 }{2}\dot{\phi} \hat{\sigma}_y,
\end{align}
i.e., it is induced by the magnetic field  $B_{y,ct}=-\dot{\phi}/2$  directed along the $y$-axis. The boundary conditions in terms of the angle $\phi$ read:
\begin{align}\label{bc}
\phi(-\infty)=-\frac{\pi}{2},\quad \phi(0)=\frac{\pi}{2}.
\end{align}
We will also assume that the spin rotates counterclockwise.
The cost function (\ref{new_cost_2}) now has the Lagrangian
\begin{align}\label{LBphi_2}
\mathcal{L}=\frac{1}{4}\dot{\phi}^2+B^2+2B\varepsilon \sin\phi+\varepsilon^2+A\cos^2\phi.
\end{align}
Since the Lagrangian \eqref{LBphi_2} does not depend explicitly on $\dot{B}$, variation over $B$ produces an algebraic constraint:
\begin{align}\label{lag-short}
\frac{\partial\mathcal{L}}{\partial B}=2B+2\varepsilon \sin\phi=0,
\end{align}
from which we find
\begin{align}\label{Bvarepsilon}
B=-\varepsilon \sin\phi.
\end{align}
At this point, we encounter a complication. By definition,  $B$ is the absolute value of the field vector, so it is always non-negative. Hence,
constraint (\ref{Bvarepsilon}) can be satisfied only for $\sin\phi \le 0$.
 Since negative values of $B$ are impossible and larger positive values of $B$ are unfavored at  $\sin\phi>0$, we have to construct our protocol so that at $\phi \le 0$ condition (\ref{Bvarepsilon}) is satisfied and at $\phi>0$ we just set $B=0$.
Let us now consider these two stages of the protocol separately.

{\bf Case $\boldsymbol{-\pi/2\le \phi<0}$}:
In this region, the field is given by Eq.~\eqref{Bvarepsilon}. Substituting \eqref{Bvarepsilon} into \eqref{LBphi_2}, we obtain
\begin{align}\label{L_model3_left}
\mathcal{L}_{\phi<0}=\frac{\dot{\phi}^2}{4}+\varepsilon^2\cos^2\phi+A\cos^2\phi.
\end{align}
This is a Lagrangian of the classical motion of a particle with mass $m=2$ in a potential $V(\phi) = -\varepsilon^2\cos^2\phi-A\cos^2\phi$.
Corresponding energy conservation equation reads:
\begin{align}\label{H_model3_left}
\frac{\dot{\phi}^2}{4}-(\varepsilon^2+A)\cos^2\phi=e.
\end{align}
Energy $e$ is found here from the observation that  at $t\rar -\infty$ we must have  $\dot{\phi} \rar 0$, and consequently $e=0$. So,
\begin{align}\label{}
\dot{\phi}=2\sqrt{\varepsilon^2+A}\cos\phi,
\end{align}
From which we find
\begin{align}\label{phi-lz}
&\phi(t)=2\arctan e^{2\sqrt{\varepsilon^2+A}(t+T)}-\frac{\pi}{2}\nn\\
&=\arctan\left\{\sinh\left[2\sqrt{\varepsilon^2+A}(t+T)\right]\right\}, \quad \phi<0.
\end{align}
At this point, we treat $T$ as a free parameter, whose value we will obtain self-consistently later.

\begin{figure}[!htb]
\scalebox{0.5}[0.5]{\includegraphics{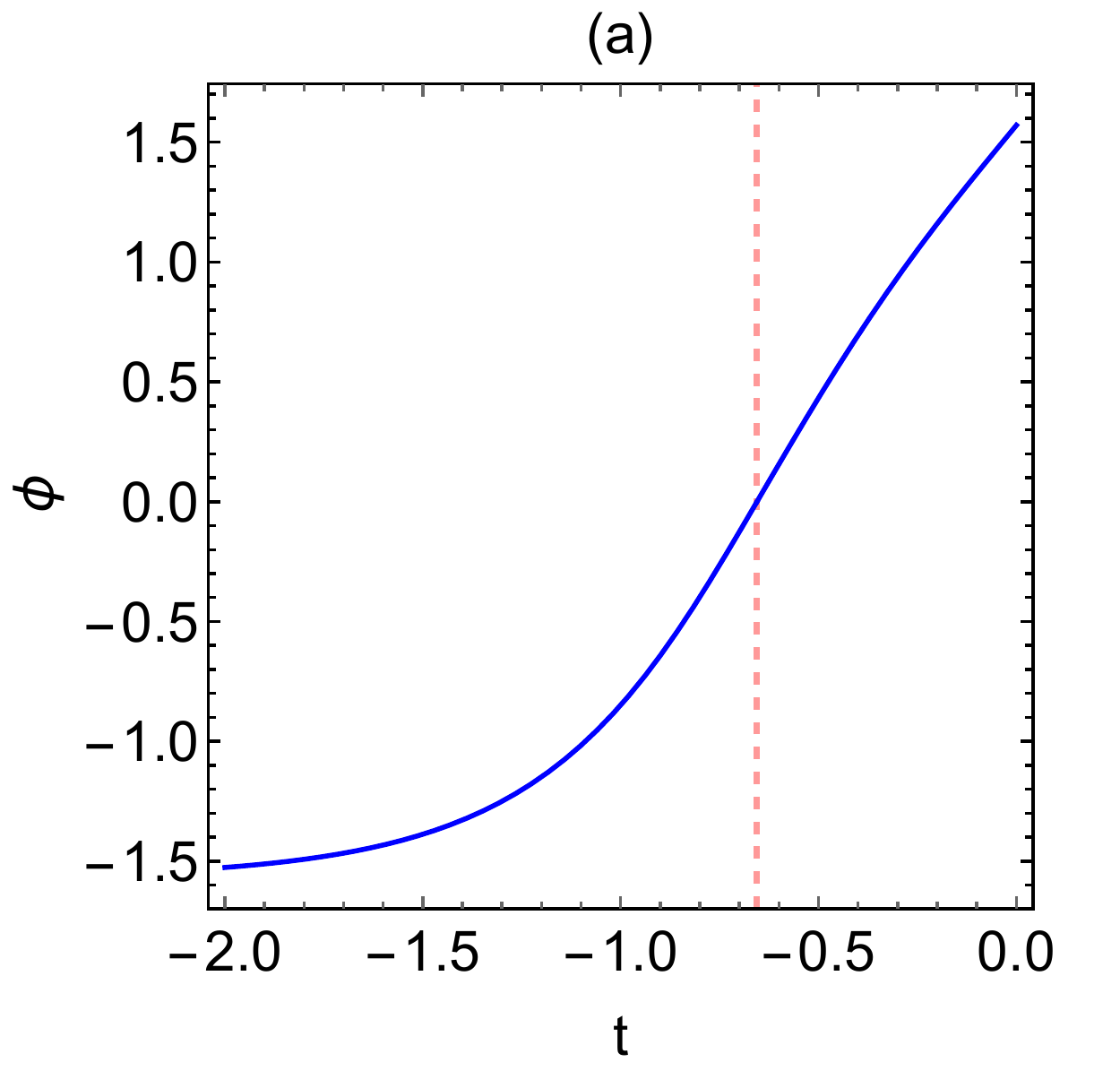}}\\
\scalebox{0.48}[0.48]{\includegraphics{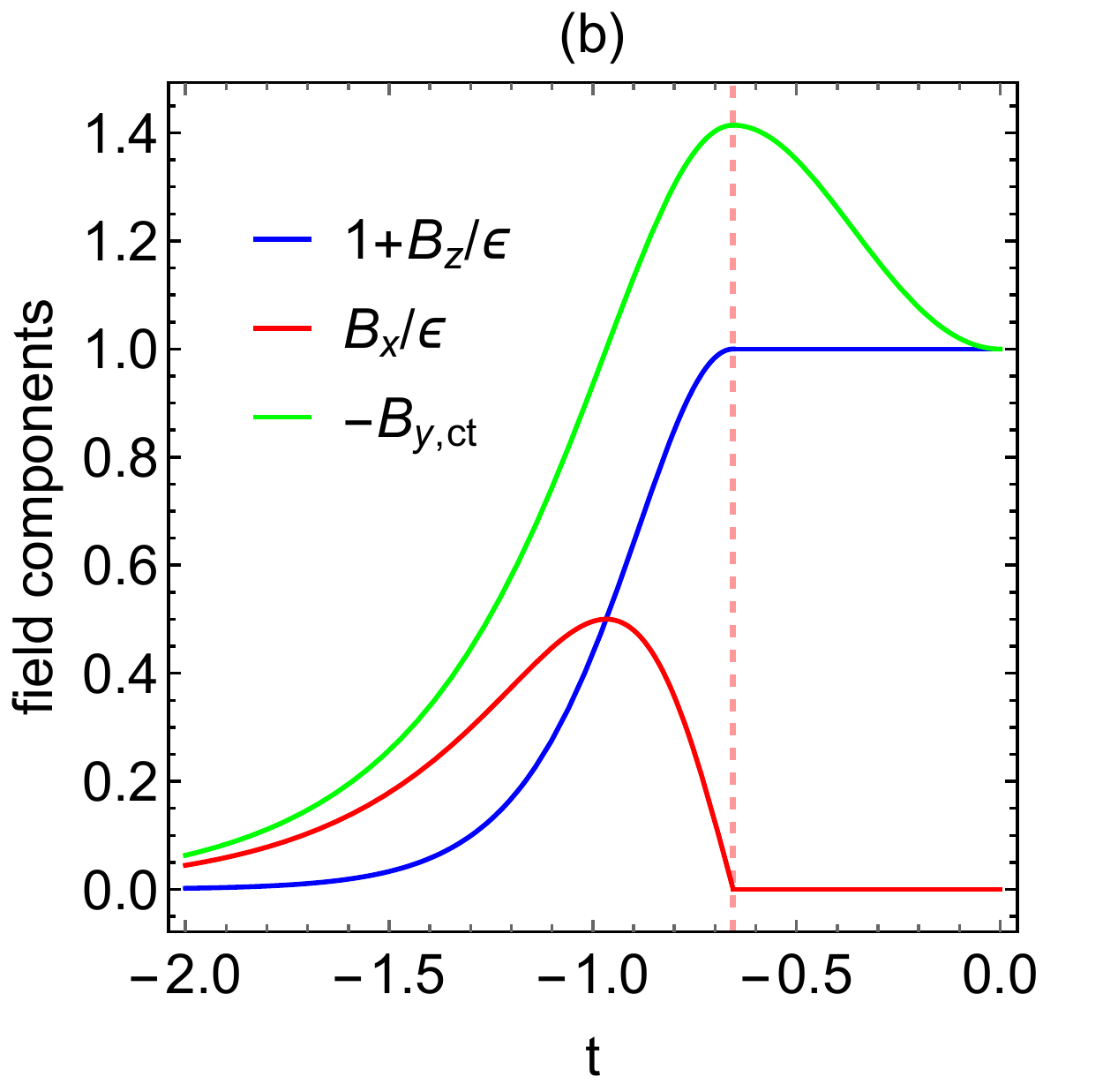}}
%\hspace{-2mm}\vspace{-4mm}%!!
\caption{The angle $\phi$ and the corresponding components of fields vs. time $t$ for the protocol described by Eqs.~(\ref{phi-lz}) and (\ref{phi-2}), at $A=1$ and $\varepsilon=1$. (a) $\phi$ vs. $t$. (b) % $-B_{y,ct}$, or $\phi'/2$, vs. $t$. (c) $B_x$ and $B_z$
The components of the applied field vs. $t$. The vertical dashed lines at $t=-T$ correspond to $\phi=0$, which is the boundary between the two cases we considered.}
\label{fig5}
\end{figure}

The field components $B_x$, $B_z$, and $B_{y,ct}$ are found as
\begin{align}
&B_x=B\cos\phi=-\varepsilon\sin\phi\cos\phi=-\varepsilon\frac{ a}{1+a^2},\label{3_soliton_bx}\\
&B_z=B\sin\phi=-\varepsilon\sin^2\phi=-\varepsilon\frac{a^2}{1+a^2},\label{3_soliton_bz}\\
&B_{y,ct}=-\dot{\phi}/2=-\frac{\sqrt{\varepsilon^2+A}}{\sqrt{1+a^2}},\\
&a\equiv\sinh\left[\sqrt{\varepsilon^2+A}\left(t+T\right)\right].\nn
\end{align}
We plot them in Fig.~\ref{fig5}(b). %(Note that $\sin\phi=\cos(2\arctan a)=\frac{1-a^2}{1+a^2}$, $\cos\phi=-\sin(2\arctan a)=-\frac{2a}{1+a^2}$.)
We can also calculate the  cost of rotating the spin from $\phi=-\pi/2$ to $\phi=0$ along this protocol:
\begin{align}
&C_{\phi<0}=\int_{-\infty}^{-T}\mathcal{L}_{\phi<0} dt=\frac{1}{2}\int_{-\infty}^{-T}\dot{\phi}^2dt=\sqrt{ \varepsilon ^2+A}.
\label{costb1}
\end{align}

{\bf Case $\boldsymbol{\pi/2 \ge \phi>0}$}:
For such values of the rotation angle we have $B=0$, and the Lagrangian \eqref{LBphi_2} reads
\begin{align}\label{L_right}
\mathcal{L}_{\phi>0}=\frac{1}{4}\dot{\phi}^2+\varepsilon^2+A\cos^2\phi.
\end{align}
This Lagrangian is the same as the one that we studied in Sec.~\ref{sine-G}. The only difference is that starting boundary conditions are now  $\phi(t=-T)=0$.
%again corresponds to the sine-Gordon equation. Its correspondng energy conservation equation reads:
%\begin{align}\label{}
%\frac{1}{4}\dot{\phi}^2-\varepsilon^2-A\cos^2\phi=E.
%\end{align}
%Thus,
%This right region is finite and $E$ is not necessarily zero. We will then treat $E$ to be a parameter, with respect to which we minimize the cost in this region. From Eq. \eqref{L_right} we have
%\begin{align}\label{}
%\dot{\phi}=2\sqrt{\varepsilon^2+A\cos^2\phi+E}.
%\end{align}
%Integrating and using the boundary conditions $\phi(t=-T)=0$ and $\phi(t=0)=\pi/2$, we obtain the time $T$ in terms of $A$, $\varepsilon$ and $E$ as:
Using results of previous section and the symmetry of evolution with Lagrangian (\ref{L_right}), we find that
\begin{align}\label{}
T=\frac{T_0}{2}=\frac{1}{2\sqrt{\varepsilon^2+A}}K\left(\frac{A}{\varepsilon^2+A}\right),
\end{align}
\begin{align}\label{phi-2}
&\phi(t)=\arcsin\left[\sn\left(2\sqrt{\varepsilon^2+A}(t+T)|\frac{A}{\varepsilon^2+A}\right)\right], \quad \phi>0,
\end{align}
and the cost of the 2nd part of the protocol:
\begin{align}
C_{\phi>0}=\sqrt{\varepsilon^2+A}E\left(\frac{A}{\varepsilon^2+A}\right).
\end{align}

Finally, we combine two costs to find
\begin{align}\label{costf}
&C^{\rm min}=C_{\phi >0}+C_{\phi<0}=\sqrt{ \varepsilon^2+A}+\sqrt{\varepsilon^2+A}E\left(\frac{A}{\varepsilon^2+A}\right).
\end{align}
Figure~\ref{fig5} summarizes dependence of the angle $\phi$ on time as given by Eqs.~(\ref{phi-lz}) and (\ref{phi-2}), as well as the components of the field that induce this spin rotation.

\begin{figure}[!htb]
\scalebox{0.5}[0.5]{\includegraphics{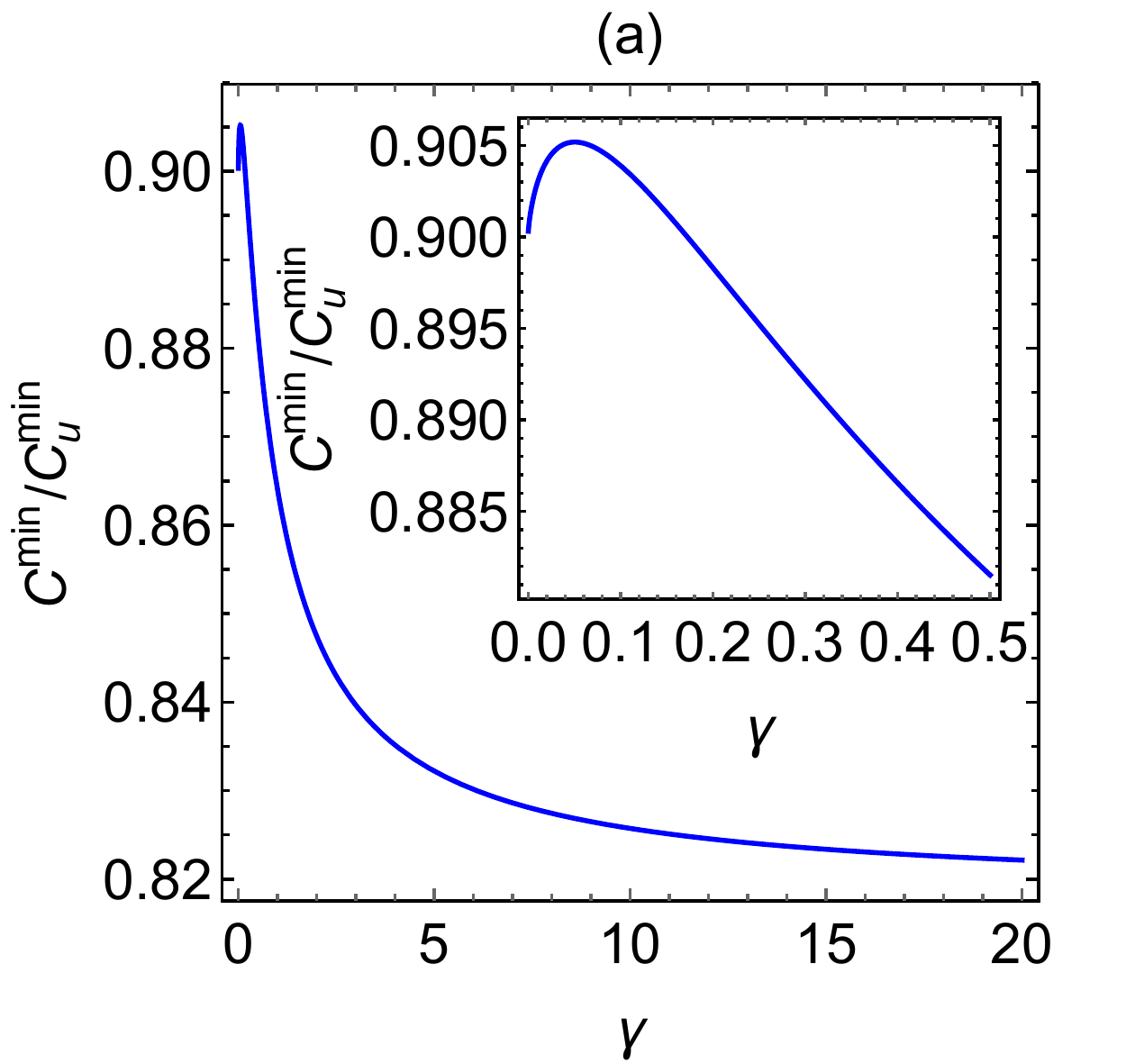}}\\
\scalebox{0.5}[0.5]{\includegraphics{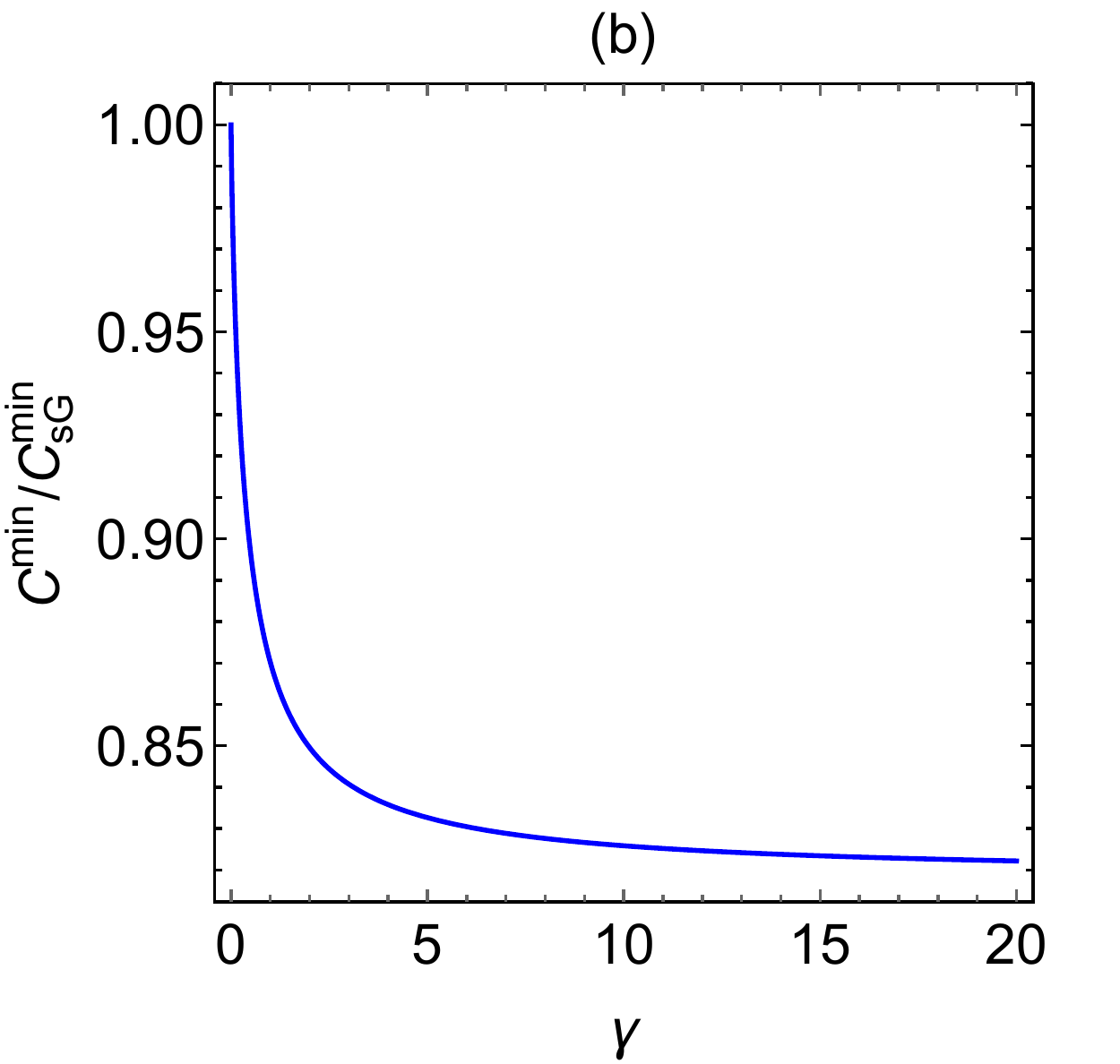}}
\caption{Comparison of costs of different protocols: (a) $C^{\rm min}/C^{\rm min}_u$ and (b) $C^{\rm min}/C^{\rm min}_{sG}$ vs. $\gamma$ ($\gamma=\varepsilon^2/A$), from $\gamma=0$ to $\gamma=20$. The inset in (a) shows a region where $\gamma$ is small: $0<\gamma<0.5$. The protocol that uses shortcuts to adiabaticity always has a smaller cost.}
\label{fig6}
\end{figure}

Let us now compare costs of optimal protocols in Eq.~(\ref{result_URP}) for uniform rotation, Eq.~(\ref{costG}) for fully compensated bias $B_z^{\rm ex}=\varepsilon$ but flexible time-dependence of $B_y(t)$, and the cost (\ref{costf}) of the protocol that used shortcuts to adiabaticity.
%with that of the uniform rotation protocol
%\begin{align}\label{}
%&C_u=\pi\sqrt{\varepsilon^2+\frac{A}{2}}.
%\end{align}
In terms of the dimensionless ratio $\gamma=\varepsilon^2/A$, these costs can be written as
\begin{align}\label{}
&C^{\rm min}=\sqrt A\sqrt{ \gamma+1}\left[1+E\left(\frac{1}{1+\gamma}\right)\right],\\
&C_{sG}^{\rm min}=2\sqrt A\sqrt{ \gamma+1}E\left(\frac{1}{1+\gamma}\right), \\
&C_u^{\rm min}=\pi\sqrt{A}\sqrt{\gamma+\frac{1}{2}}.
\end{align}
We find that ratios of these costs depend only on $\gamma$. Figure~\ref{fig6} presents the ratios $C^{\rm min}/C_u^{\rm min}$ and $C^{\rm min}/C_{sG}^{\rm min}$ vs. $\gamma$. It shows that these two ratios are smaller than 1 for any $\gamma$. Thus, for any choices of parameters $A$ and $\varepsilon$, the protocol using shortcuts to adiabaticity has a lower cost than the protocols based on instantaneous bias compensation. For small and large $\gamma$, the limits of the two ratios are
\begin{align}\label{}
&\operatorname{lim} \limits_{\gamma\rar 0}\frac{C^{\rm min}}{C^{\rm min}_u}=\frac{2\sqrt{2}}{\pi}\approx0.900,\label{gamma0_2}\\
&\operatorname{lim} \limits_{\gamma\rar\infty}\frac{C^{\rm min}}{C^{\rm min}_u}=\frac{1}{2}+\frac{1}{\pi}\approx0.818,\label{gammainfty_2}\\
&\operatorname{lim} \limits_{\gamma\rar 0}\frac{C^{\rm min}}{C^{\rm min}_{sG}}=1,\\
&\operatorname{lim} \limits_{\gamma\rar\infty}\frac{C^{\rm min}}{C^{\rm min}_{sG}}=\frac{1}{2}+\frac{1}{\pi}\approx0.818.
\end{align}
Between the two limits, $C^{\rm min}/C^{\rm min}_u$ has a maximum at $\gamma=0.0504$, %$\gamma=0.05036$
with $C^{\rm min}/C^{\rm min}_u=0.905$. %$C/C_u=0.09052$.
Thus, for any $A$ and $\varepsilon$ the protocol that uses shortcuts to adiabaticity leads up to $18.2$ percent reduction of the cost compared to the protocols based on instantaneous compensation of the natural bias term in the Hamiltonian. %Note that in the $\gamma\rar \infty$, or $\varepsilon^2\gg A $ limit, our protocol in the right region approaches  the uniform rotation protocol, and this contributes the term $1/2$ to $C/C_u$ in Eq. \eqref{gammainfty_2}.
%The fully compensated bias protocol has an intermediate performance among the three protocols.

\section{Conclusions}

We demonstrated the possibility of a mostly analytical approach to quantum control problems.
Instead of  pursuing for numerically exact but complex and nontransparent solutions we explored the possibility  to restrict the optimization problem to some sufficiently broad classes of analytically tractable protocols.
By construction, such protocols are generally suboptimal.  The precise cost function is, however, unknown in most practical situations.
There is no reason then to know the numerically exact solution of an optimization problem. It is much more desirable for practically useful protocols to  be physically meaningful and analytically tractable.
Such a control can be relatively easily planned and adjusted upon receiving additional information about performance. This simplicity is what our approach can provide.

By restricting the search for optimal protocols to a class of solutions of some exactly solvable model, we automatically resolve the problem of combinatorial complexity of quantum systems  because state vector components no longer have to be treated as independent parameters. As an additional bonus, such a restriction leads to  control field pulses whose shapes can be written in terms of known special functions, and consequently can be easily characterized. Numerical solution may be needed only for a small number of saddle point equations for control parameters,

Using the spin-1/2 example, we demonstrated two basic possibilities for choosing the family of solutions for optimization. One is based on simple compensation of terms in the Hamiltonian that prevented an analytical solution to exist in the first place. Another approach is based on using families of solutions obtained by solving an inverse problem, i.e., assuming some functional form of the state vector evolution  (a shortcut to adiabaticity in our case) and then finding the family of Hamiltonians that produces this evolution.
By restricting the search for optimal control among such classes of analytically solvable Sch\"odinger equations, we were able to express the cost functional  only in terms of control parameters.

The worked out example revealed not only the simplicity of our approach but also potential problems that may become important in more complicated situations. 
For example, protocols that are based on straightforward removal of unwanted couplings  are simple and generally need only control of the local degrees of freedom. We showed, however, that more complex protocols can be substantially less costly. We also found that,  interestingly, it was impossible to build the optimal protocol fully within the family of  shortcuts to adiabaticity because equations of motion drove one of the parameters beyond its range of definition. We resolved this problem by combining different types of protocols in order to describe different time intervals. Resulting protocol, however, had the best performance among all strategies that we explored. It is expected that similar properties of optimized control protocols  will be generally found in multistate problems too. So, it seems that more complex models will still require such manual resolution of complications.  

Finally, we note that we left many characteristics of our approach unstudied. For example, in addition to extending it to more complex quantum problems, it should be useful to explore the robustness of our protocols with respect to uncertainty of parameters, e.g., using the integral nullification method introduced in \cite{Daems}. It should be also interesting to compare our protocols with ones based on more complex, than shortcuts to adiabaticity, classes of solvable models that can be derived by inverting the Schr\"odinger equation \cite{garanin}.

\section*{Acknowledgements}
The work was carried out under the auspices of the National Nuclear Security Administration of the U.S. Department of Energy at Los
Alamos National Laboratory under Contract No. DE-AC52-06NA25396. Authors also thank the support from the LDRD program at LANL.

\section*{Appendix: Elliptic integrals and Jacobi elliptic functions}
Here we list special functions that we have used in this text, namely, the elliptic integrals and Jacobi elliptic functions.

The incomplete elliptic integral of the first kind, $F(\phi|m)$, with modulus $m$ is defined as
\begin{align}\label{def_F}
&F(\phi|m)=\int_0^{\phi}d\theta\frac{1}{\sqrt{1-m\sin^2\theta}}.
\end{align}
The complete elliptic integral of the first kind, $K(m)$, is defined as
\begin{align}\label{def_K}
&K(m)=F\left(\frac{\pi}{2}|m\right)=\int_0^{\frac{\pi}{2}}d\phi\frac{1}{\sqrt{1-m\sin^2\phi}}.
\end{align}
The incomplete elliptic integral of the second kind, $E(\phi|m)$, is defined as
\begin{align}\label{def_E}
&E(\phi|m)=\int_0^{\phi}d\theta\sqrt{1-m\sin^2\theta},
\end{align}
and the complete elliptic integral of the second kind, $E(m)$, is defined as
\begin{align}\label{def_Ec}
&E(m)=E\left(\frac{\pi}{2}|m\right)=\int_0^{\frac{\pi}{2}}d\phi\sqrt{1-m\sin^2\phi}.
\end{align}

The Jacobi elliptic functions are the inverses of the incomplete elliptic integrals of the first kind \cite{wiki-Jef}. If $u=F(\phi|m)$, then the Jacobi elliptic function $\operatorname{sn}$(x|m)  %\operatorname{cn}, \operatorname{dn}
is defined as \cite{wiki-Jef}:
\begin{align}\label{}
&\sn (u|m)=\sin\phi.\label{def_sn}
%&\cn (u|m)=\cos\phi,\label{def_cn}\\
%&\dn (u|m)=\sqrt{1-m\sin^2\phi}.\label{def_dn}
\end{align}

\begin{comment}
The series expansions of $\sn$, $\cn$ and $\dn$ at $m\rar 0$ read \cite{DLMF}:
\begin{align}\label{}
&\sn(u|m)=\sin u-\frac{m}{4}[(u-\sin u\cos u)\cos u]+O(m^2),\label{series_sn_0}\\
&\cn(u|m)=\cos u+\frac{m}{4}[(u-\sin u\cos u)\sin u]+O(m^2),\label{series_cn_0}\\
&\dn(u|m)=1-\frac{m}{2}\sin^2 u+O(m^2),\label{series_dn_0}
\end{align}
while the series expansions at $m\rar 1$ read \cite{DLMF}:
\begin{align}\label{}
&\sn(u|m)=\tanh u-\frac{1-m}{4}(u-\sinh u\cosh u)\sech^2 u\nn\\
&+O((1-m)^2),\label{series_sn_1}\\
&\cn(u|m)=\sech u+\frac{1-m}{4}(u-\sinh u\cosh u)\tanh u \sech u\nn\\
&+O((1-m)^2),\label{series_cn_1}\\
&\dn(u|m)=\sech u+\frac{1-m}{4}(u+\sinh u\cosh u)\tanh u \sech u\nn\\
&+O((1-m)^2).\label{series_dn_1}
\end{align}
\end{comment}

\end{document}